\documentstyle[graphicx,aps]{revtex}
\begin{document}
\title{Point defects in Hard Sphere Crystals}
\author{Sander Pronk and Daan Frenkel\footnote{e--mail: frenkel@amolf.nl}}
\address{FOM Institute for Atomic and Molecular Physics\\ 
Kruislaan 407\\
1098 SJ Amsterdam\\
the Netherlands}
\maketitle

\begin{abstract}
We report numerical calculations of the concentration of interstitials in
hard-sphere crystals. We find that, in a three-dimensional fcc hard-sphere
crystal at the melting point, the concentration of interstitials is $%
2.7(4)\cdot 10^{-8}$. This is some three orders of magnitude lower than the
concentration of vacancies. A simple, analytical estimate yields a value
that is in fair agreement with the numerical results. 
\end{abstract}

\section{Introduction}

Any crystal in equilibrium will contain defects, such as vacancies,
interstitials and dislocations. Of these, the point defects are the most
common. Some thirty years ago, Bennett and Alder\cite{BennettAlder}
estimated the equilibrium concentration of vacancies in a hard-sphere
crystal and found that, close to melting, this concentration could be quite
high (typically, one vacancy per $4000$ lattice sites). At present, the
question of the concentration (and transport) of point defects in
(colloidal) hard-sphere crystals takes on a renewed --- and now quite
practical --- significance. Apart from the theoretical interest in
hard sphere crystals as a model system, crystals from colloidal particles,
having lattice sizes comparable to the wavelength of light, are being 
prepared and studied because of their potentially interesting photonic 
properties. Clearly the
presence of even a small number of defects can have a pronounced effect on
the nature of photonic states in such materials. Moreover, as the accuracy
of free energy calculations increases, it is no longer permissible to ignore
the contribution of vacancies to the total free energy. The aim of the
present paper is to review briefly the statistical mechanical description of
a crystal with point defects. This problem is not completely trivial, as the
concept of a vacancy or interstitial is inextricably linked to that of
lattice sites. And lattice sites loose their meaning in a disordered state.
So, we should first address the question: when is it permissible to count
states with a different number of lattice sites as distinct? The answer is,
of course, that this is only true if these different states can be assigned
to distinct volumes in phase space. This is possible if we impose that every
particle in a crystal is confined to its Wigner-Seitz cell. In three
dimensional crystals, this constraint on the positions of all particles has
little effect on the free energy (in contrast, in a {\em liquid} it is not
at all permissible). In a two-dimensional crystal, the constraint is more
problematic, at least in the thermodynamic limit. However, for large but
finite two-dimensional systems, the single-occupancy cell constraint is also
quite reasonable. Below, we describe two alternative (but equivalent) routes
to arrive at the free energy of a crystal with vacancies. In one case, we
use the Grand-Canonical ensemble. This would seems to be the most obvious 
ensemble to use when describing a system with a fluctuating number of 
particles. Yet, the analysis is complicated by the fact that not only the 
number of particles, but also the number of lattice sites, may fluctuate. 
In the second analysis, we consider an isothermal-isobaric system. The latter
approach is simpler and is, apart from a minor correction, equivalent to the
one followed by Bennett and Alder\cite{BennettAlder}. We then describe our
numerical approach to compute the concentration of interstitials in a
hard-sphere crystal. We compare our numerical results with a simple
theoretical estimate. 

\section{Free energy of vacancies}

\subsection{The grand-canonical route}

When considering the statistical mechanics of a crystal with vacancies, it
is convenient to consider first a system with a fixed number of lattice
sites, $M$, contained in a volume $V$. If this crystal is in contact with a
particle-reservoir at chemical potential $\mu$, then the number of vacancies
in the crystal may fluctuate. In principle, the crystal could also contain
interstitials but, for the time being, we shall ignore this possibility. It
is then easy to write down the expression for the grand potential of the
crystal $\Xi^{\prime}$: 
\begin{equation}
\Xi_{M}^{\prime}=\sum_{n=0}^{M}\exp \Big[(M-n)\beta\mu \Big] Q_{M-n}(V,T)
\label{Grand1}
\end{equation}
where $\beta \equiv 1/k_BT$. Note that this is not the true grand potential,
because we should also allow for fluctuations in the number of lattice
sites. We denote the free energy of a crystal with {\em no} vacancies by $%
F^{(0)}=-kT\ln Q_{M}$. In practice, the equilibrium concentration of
vacancies in a crystal is very low. We shall therefore make the
approximation that vacancies do not interact. This assumption is not as
reasonable as it seems, as the interaction of vacancies through the stress
field is quite long-ranged. The assumption that vacancies are ideal makes it
easier to compute the canonical partition function of a crystal with $n$
vacancies: 
\begin{equation}
Q_{M-n}(V,T)\approx\frac{M!}{n!(M-n)!}Q^{(n)}(V,T)=\frac{M!}{n!(M-n)!}%
\exp(-\beta F^{(n)})  \label{Qnvac}
\end{equation}
where we have used the notation $F^{(n)}$ to denote the free energy of a
crystal with $n$ vacancies {\em at given positions}. As the vacancies are
assumed to be non-interacting, it is clear that we can write 
\begin{equation}
F^{(n)}=F^{(0)}-nf_{1}=Mf_{0}-nf_{1}  \label{Fnvac2}
\end{equation}
where $f_{0}$ is the free energy per particle in the defect-free crystal and
-$f_{1}$ is the change in free energy of a crystal due to the creation of a
single vacancy at a specific lattice point\footnote{%
The choice of the minus sign in the definition will later turn out to be
convenient.}. Combining Eqns. (\ref{Grand1}), (\ref{Qnvac}) and (\ref{Fnvac2}), 
%(\ref{Grand1})--(\ref{Fnvac2}), we 
we obtain
\begin{eqnarray}
\Xi_{M}^{\prime} & = &\sum_{n=0}^{M}\frac{M!}{n!(M-n)!} \exp \Big[%
(M-n)\beta\mu \Big] \exp \Big[-\beta( Mf_{0}-nf_{1}) \Big]  \nonumber \\
& \equiv & Q_{M}\exp(M\beta\mu)\bigg[ 1+ \exp\Big(-\beta\lbrack\mu-f_{1}]%
\Big)\bigg] ^{M}  \label{fvacancy1}
\end{eqnarray}
Usually, $\exp[-\beta(\mu-f_{1})]$ is much less than unity. This allows us
to write: 
\begin{equation}
\Xi_{M}^{\prime}=Q_{M}\exp(M\beta\mu) \exp \bigg[ M\exp \Big(%
-\beta\lbrack\mu-f_{1}] \Big) \bigg]  \label{Fvacancy}
\end{equation}
Using 
\[
\left\langle M-n\right\rangle =\frac{\partial\ln\Xi^{\prime}}{\partial\beta
\mu} 
\]
the average number of vacancies follows as 
\begin{equation}
\left\langle n\right\rangle =M\exp\Big[-\beta(\mu-f_{1})\Big]  \label{Nvac}
\end{equation}
Now, we should take into account the fact that, actually, the number of
lattice sites itself is not fixed but will adjust to the number of
vacancies. The total grand partition function is therefore a sum over all
states with different number of lattice sites $M^{\prime}=M+\Delta M$. In
practice, $\Delta M\ll M$. We can then write 
\[
\Xi=\sum_{\Delta M=-\infty}^{\infty}\Xi_{M+\Delta M}^{\prime} 
\]
Note that $\Xi^{\prime}$ depends on both $M+\Delta M$ and $\mu$. We now
choose the reference number of lattice sites such that in that particular
case $\mu$ is equal to the chemical potential of the perfect lattice. That
is: 
\[
F^{(0)}+P^{(0)}V=M\mu 
\]
We also introduce a rather strange quantity, namely the ``grand canonical''
partition function of a perfect lattice with a fixed number of particles $%
(M):$ $\Xi_{M}^{(0)}\equiv Q_{M}\exp(M\beta\mu)$. Of course, $\Xi_{M}^{(0)}$
is {\em not} a true grand-canonical partition function, as the number of
particles in this system is fixed. The Grand Canonical partition function
then becomes: 
\begin{eqnarray}
\Xi & = & \sum_{ \Delta M=-\infty}^{\infty} \exp \Big[-\beta
F^{(0)}(M+\Delta M) \Big]  \nonumber \\
& \times & \exp \Big[(M+\Delta M)\beta\mu \Big] \exp \bigg[\Big(M+\Delta M%
\Big)\exp \Big(-\beta\lbrack \mu-f_{1}] \Big) \bigg]  \label{Xi1}
\end{eqnarray}
We assume (as usual) that, in the thermodynamic limit, $\Xi$ is dominated by
the largest term in the sum. Hence, we have to determine the point where the
derivative of $\Xi$ with respect to $M$ vanishes. To this end, we perform a
Taylor expansion of the exponent in Eqn. (\ref{Xi1}) in powers of 
$\Delta M$. Note that for a perfect --- defect free --- lattice 
\begin{equation}
\frac{\partial F^{(0)}}{\partial M}=\frac{\partial F^{(0)}}{\partial N}=\mu
\label{mudef}
\end{equation}
and 
\begin{equation}
\frac{\partial^{2}F^{(0)}}{\partial M^{2}}=\frac{\partial\mu}{\partial N}=%
\frac{1}{N}\frac{\partial P}{\partial\rho}
\end{equation}
Moreover, 
\begin{equation}
f_{1}(M+\Delta M)=f_{1}(M)+\frac{1}{V}\frac{\partial f_{1}}{\partial\rho }%
\Delta M+{\cal O}(\Delta M^{2})  \label{ftaylor}
\end{equation}
By combining Eqn. (\ref{mudef})--(\ref{ftaylor}) we obtain 
\begin{equation}
\exp\Big[-\beta(\mu-f_{1}[M+\Delta M])\Big]=\exp\Big\{-\beta[\mu-f_{1}(M)]%
\Big\} \times \Big[1+\frac{\beta}{V}\frac{\partial f_{1}}{\partial\rho}%
\Delta M+{\cal O}(\Delta M^{2})\Big]
\end{equation}
Note that $f_{1}=F^{(0)}-F^{(1)}$. Hence, 
\begin{equation}
\frac{\beta}{V}\frac{\partial f_{1}}{\partial\rho}=\frac{\beta}{V}\frac{%
\partial f_{1}}{\partial V}\frac{\partial V}{\partial\rho}=\beta\frac {V}{N}%
\left( P^{(0)}-P^{(1)}\right) \equiv\beta\Delta P^{0,1}/\rho
\label{DeltaP01}
\end{equation}
where $\Delta P^{0,1}$ is the difference in the pressure of two crystals
with $M$ lattice sites, one with zero vacancies and the other with one
vacancy (both in a fixed volume $V$ and a temperature $T$). As a
consequence, 
\begin{eqnarray}
& (M+\Delta M) &\exp \Big[-\beta(\mu-f_{1}[M+\Delta M]) \Big]  \nonumber \\
& & \approx M\exp \Big\{-\beta[\mu-f_{1}(M)] \Big\} \bigg\{1+\frac{\Delta M}{%
M}+(\beta\Delta P^{0,1}/\rho)\Delta M\bigg\}
\end{eqnarray}
Inserting the Taylor expansion in the expression for $\Xi$, we obtain 
\begin{eqnarray}
\Xi = & \exp \Big[-\beta F^{(0)}(M)+M\beta\mu \Big]\sum_{\Delta M} \exp%
\bigg\{M\exp \Big[-\beta(\mu-f_{1}(M)) \Big]  \nonumber \\
& \times\Big[1+\frac{\Delta M}{M}+ (\beta\Delta P^{0,1}/\rho)\Delta M\Big]-%
\frac{\beta}{2}\frac{1}{M}\frac{\partial P}{\partial\rho}(\Delta M)^{2}%
\bigg\}  \label{Xiexp}
\end{eqnarray}
It may seem inconsistent that we expand to second order in $\Delta M$ in the
last term of Eqn. (\ref{Xiexp}) but only to first order in the preceding
terms. However, as we show below, we actually expand --- consistently --- to
second order in the vacancy concentration.

We define the fractional change in the number of lattice sites, $y$, as 
\begin{equation}
y\equiv\frac{\Delta M}{M}
\end{equation}
Similarly, we define the vacancy concentration $x$ as 
\begin{equation}
x=\frac{\left\langle n\right\rangle }{M}= \exp \Big[-\beta(\mu-f_{1}) \Big]
\end{equation}
Finding the maximum in the exponent is then equivalent to maximizing 
\begin{equation}
-\frac{\beta M}{2}\left( \frac{\partial P}{\partial\rho}\right) y^{2}+Mx(0)%
\Big[1+y(1+\beta V\Delta P^{0,1})\Big]
\end{equation}
where $x(0)$ is the value of $x$ for $\Delta M=0$ and $\rho_{s}$ is the
density of the ideal reference lattice. The maximum value is at 
\begin{equation}
y=x(0)\frac{(1+\beta V\Delta P^{0,1})}{\beta\left( \frac{\partial P}{%
\partial\rho}\right) }
\end{equation}
and the maximum of the grand canonical potential is 
\begin{equation}
\Xi\approx\exp(-\beta F^{(0)}(M)+M\beta\mu)\exp\left( Mx(0)+\frac{M}{2}\frac{%
x^{2}(0) \Big(1+\beta V\Delta P^{0,1} \Big)^{2}}{\beta\left( \frac{\partial P%
}{\partial\rho}\right) }\right)  \label{XiTotal}
\end{equation}
Hence, the presence of vacancies increases the grand potential (as it
should) {\em and it changes\ (increases) the number of lattice sites}.
However, ignoring terms of order ${\cal O}(x^{2}(0))$, the vacancy
concentration is still given by Eqn. (\ref{Nvac}).

The next question is: how do vacancies affect the melting curve. Now our
definition of the reference system (i.e. the perfect lattice with the same $%
\mu$) turns out to be convenient. Note that the Grand Canonical partition
function is related to the pressure by 
\begin{equation}
\beta PV=\ln\Xi=\beta P^{(0)}V+Mx(0)+{\cal O}(x(0)^{2})
\end{equation}
Ignoring terms quadratic in the vacancy concentration, we find that the
effect of allowing for vacancies is to increase the pressure of the solid by
an amount 
\begin{equation}
{\bf \Delta P}\approx x(0)\rho_{s}kT  \label{deltaPsolid}
\end{equation}
Let us assume that the liquid was in equilibrium with the perfect crystal at
pressure $P$ and chemical potential $\mu$. Then it is easy to verify that
the shift in the coexistence pressure due to the presence of vacancies is 
\begin{equation}
%\delta P_{coex}=\frac{\rho_{s}{\bf \Delta P}}{\rho_{s}-\rho_{l}}
\delta P_{coex}=\frac{-x(0)kT}{\rho_{l}^{-1}-\rho_{s}^{-1}}
\end{equation}
and the corresponding shift in the chemical potential at coexistence is 
\begin{equation}
\delta\mu_{coex}=\frac{\delta P_{coex}}{\rho_{l}}
\end{equation}
Direct calculations of the vacancy concentration in a hard-sphere crystal at
melting \cite{BennettAlder} indicate that $x(0)\approx2.6 \cdot 10^{-4}$.
Hence, the increase in the coexistence pressure due to vacancies is 
%$\delta
%P_{coex}\approx2.68 \cdot 10^{-3}$. The corresponding shift in the chemical
%potential at coexistence is $\delta\mu_{coex}=2.86 \cdot 10^{-3}$. 
$\delta P_{coex}\approx -2.57 \cdot 10^{-3}$ $kT/\sigma^3$ 
(where $\sigma$ is the particle diameter). 
The corresponding shift in the chemical
potential at coexistence is $\delta \mu _{coex}=-2.74 \cdot 10^{-3} kT$
Note that
this shift is very significant when compared to the accuracy of absolute
free-energy calculations of the crystalline solid\cite{Polson}.

\subsection{The NPT route}

Bennett and Alder\cite{BennettAlder} work with the Gibbs free energy rather
than the Helmholtz free energy. Their expression for the vacancy
concentration is based on the analysis of the effect of vacancies on the
Gibbs free energy of a system of $N$ particles at constant pressure and
temperature. First, we define $g^{vac}$, the variation in the Gibbs free
energy of a crystal of $M$ particles due to the introduction of a single
vacancy {\em at a given lattice position} 
\begin{eqnarray}
g^{vac} & \equiv & G_{M+1,1}(N,P,T)-G_{M,0}(N,P,T)  \nonumber \\
& = &F_{M+1,1}(V_{M+1,1})-F_{M,0}(V_{M,0})+P(V_{M+1,1}-V_{M,0})
\end{eqnarray}
where the first subscript refers to the number of lattice sites in the
system, and the second subscript to the number of vacancies. In this
equation we distinguish $M$, the original number of lattice sites, and $N$,
the number of particles, even though in the present case $N=M.$ Let us write 
$f_{1}$ (Eqn. (\ref{f1})) as 
\[
-f_{1}\equiv F_{M+1,1}(V_{M+1,0})-F_{M+1,0}(V_{M+1,0}) 
\]
Hence, 
\begin{eqnarray}
g^{vac} & = & F_{M+1,1}(V_{M+1,1})-F_{M+1,1}(V_{M+1,0})+  \nonumber \\
& & F_{M+1,1}(V_{M+1,0})-F_{M+1,0}(V_{M+1,0})+  \nonumber \\
& & F_{M+1,0}(V_{M+1,0})-F_{M,0}(V_{M,0})+  \nonumber \\
& & P(V_{M+1,1}-V_{M,0})
\end{eqnarray}
The next step is to introduce a hypothetical defect free crystal with $M$
lattice sites, at the same pressure as the system with $M+1$ lattice sites.
The volume of this system is $V_{M,0}=\{M/(M+1)\}V_{M+1,0}.$Similarly, the
free energy of this hypothetical system is $F_{M,0}=\{M/(M+1)\}F_{M+1,0}.$
Note also that 
\begin{equation}
F_{M+1,1}(V_{M+1,1})-F_{M,0}(V_{M,0})=-P\Delta v-f_{1}+f_{0}
\end{equation}
where $\Delta v\equiv v^{vac}-v^{part}$ is the difference in volume of a
vacancy and a particle, at constant pressure and number of lattice sites.
Moreover, 
\begin{equation}
P(V_{M+1,1}-V_{M,0})=P(\Delta v+V/N)
\end{equation}
Hence, the Gibbs free energy difference associated with the formation of a
vacancy at a specific lattice site, $G_{M,1}-G_{M-1,0}\equiv g^{vac}$, is
then 
\begin{eqnarray}
g^{vac} & = &P(V_{M+1,1}-V_{M,0})-f_{1}+(\Delta v+V/N)P+f_{0}  \nonumber \\
& = & P(V_{M,1}-V_{M,0}+V_{M,0}-V_{M-1,0})-f_{1}+f_{0}  \nonumber \\
& = & P(V/N)-f_{1}+f_{0}  \nonumber \\
& = & (P/\rho+f_{0})-f_{1}  \nonumber \\
& = & \mu_{0}-f_{1}  \label{gvacmuf1}
\end{eqnarray}
where we have defined $\mu_{0}\equiv(P/\rho+f_{0})$. Now we have to include
the entropic contribution due to the distribution of $n$ vacancies over $M$
lattice sites. This total Gibbs free energy then becomes 
\begin{eqnarray}
G & = & G_{0}(N)+ng^{vac}+MkT\left( \frac{n}{M}\ln\frac{n}{M}+\left[1-\frac{n%
}{M}\right]\ln \left[1-\frac{n}{M}\right] \right) \\
& \approx & G_{0}(N)+ng^{vac}+nkT\ln\frac{n}{M}-nkT
\end{eqnarray}
If we minimize the Gibbs free energy with respect to $n$, we find 
\[
\left\langle n\right\rangle \approx M\exp(-\beta g^{vac}) 
\]
where we have ignored a small correction due to the variation of $\ln M$
with $n$. If we insert this value in the expression for the total Gibbs free
energy, we find: 
\[
G=G_{0}(N)+\left\langle n\right\rangle g^{vac}-\left\langle n\right\rangle
g^{vac}-\left\langle n\right\rangle kT=G_{0}-\left\langle n\right\rangle kT 
\]
The total number of particles is $M-$ $\left\langle n\right\rangle $. Hence
the Gibbs free energy {\em per particle} is 
\begin{eqnarray}
\mu & = &\frac{G_{0}-\left\langle n\right\rangle kT}{N}=\mu_{0}-\frac {%
\left\langle n\right\rangle kT}{N}  \nonumber \\
& \approx &\mu_{0}-x_{v}kT  \label{gvacfinal}
\end{eqnarray}
Thus the change in chemical potential of the solid is 
\begin{equation}
\Delta\mu=-x_{v}kT
\end{equation}
from which it follows that the change in {\em pressure} of the solid at
fixed chemical potential is equal to 
\begin{equation}
\Delta P=x_{v}\rho_{s}kT
\end{equation}
This is equivalent to Eqn. (\ref{deltaPsolid}) above. Hence the
Bennett-Alder scheme is equivalent to the ``grand-canonical'' 
scheme 
\footnote{
In Ref. \cite{BennettAlder}, a slightly different expression is found, but
this is due to a small error in the derivation in that paper.}

It should be pointed out that the variation in volume due to the replacement
of a particle by a vacancy can be computed either directly, in a
constant-pressure simulation, or indirectly by measuring the change in
pressure in a constant volume simulation. The two methods are related
through the thermodynamic relation 
\begin{equation}
\left( \frac{\partial V}{\partial P}\right) _{N,T}\left( \frac{\partial P}{%
\partial N}\right) _{V,T}\left( \frac{\partial N}{\partial V}\right)
_{P,T}=-1
\end{equation}
Noting that the number of vacancies is $n=M-N$, we can see that the change
in pressure with the number of vacancies for a fixed number of lattice
sites, is 
\begin{equation}
-\left( \frac{\partial P}{\partial N}\right) _{V,T}=\left( \frac{\partial P}{%
\partial V}\right) _{N,T}\left( \frac{\partial V}{\partial N}\right) _{P,T}
\end{equation}
In particular, the pressure change due to one vacancy (i.e. $\Delta P^{0,1}$%
, defined in Eqn. (\ref{DeltaP01})) is 
\begin{equation}
\Delta P^{0,1}=P_{M.0}-P_{M,1}=\Delta v\left( \frac{\partial P}{\partial V}%
\right) _{N,T}
\end{equation}

\section{Computational scheme}

\subsection{Vacancies}

Numerically, it is straightforward to compute the equilibrium vacancy
concentration. As before, the central quantity that needs to be computed is $%
-f_{1}$, the change in free energy of a crystal due to the creation of a
single vacancy at a specific lattice point. In fact, it is more convenient
to consider $+f_{1}$, the change in free energy due to the removal of a
vacancy at a specific lattice point. This quantity can be computed in
several ways. For instance, we could use a particle-insertion method. We
start with a crystal containing one single vacancy and attempt a trial
insertion in the Wigner-Seitz cell surrounding that vacancy. Then $f_{1}$ is
given by 
\begin{equation}
f_{1}=-kT\ln\left( \frac{V_{WS}<\exp(-\beta\Delta U>}{\Lambda^{d}}\right)
\label{f1}
\end{equation}
where $V_{WS}$ is the volume of the Wigner-Seitz cell, and $\Delta U$ is the
change in potential energy associated with the insertion of a trial
particle. For hard particles 
\begin{equation}
f_{1}=-kT\ln\left( \frac{V_{WS}P_{acc}(V_{WS})}{\Lambda^{d}}\right)
\end{equation}
where $P_{acc}(V_{WS})$ is the probability that the trial insertion in the
Wigner-Seitz cell will be accepted. As most of the Wigner-Seitz cell is not
accessible, it is more efficient to attempt insertion in a sub-volume
(typically of the order of the cell-volume in a lattice-gas model of the
solid). However, then we also should consider the reverse move --- the
removal of a particle from a sub-volume $v$ of the Wigner-Seitz cell, in a
crystal without vacancies. The only thing we need to compute in this case is 
$P_{rem}(v)$, the probability that a particle happens to be inside this
volume. The expression for $f_{1}$ is then 
\begin{equation}
f_{1}=-kT\ln\left( \frac{vP_{acc}(v)}{P_{rem}(v)\Lambda^{d}}\right) \text{ .}
\end{equation}
Of course, in the final expression for the vacancy concentration, the factor 
$\Lambda^{d}$ drops out (as it should), because it is cancelled by the same
term in the ideal part of the chemical potential.

\subsection{Interstitials}

As in the case of vacancies, the calculation of interstitials centers around
the calculation of $g_I$, the free energy associated with introducing an
interstitial into the system (in the NPT ensemble). $g_I$ can be expressed
as a sum of two parts: the free energy of introducing a point particle into
the system ($g^{ins} = -kT \ln <1 - \eta>$, where $\eta$ is the packing
fraction), and the free energy of growing that particle to the same diameter
($\sigma_0 \equiv 1$) as the other particles ($g^{grow}$).

For the calculation of $g^{grow}$, we simulate an extended system consisting
of a lattice of $N$ particles with diameter $\sigma _{0}=1$ on (or near)
lattice sites and one extra (interstitial) particle that has a diameter $%
\sigma _{I}$ that can vary freely. We interpret $\sigma _{I}$ as an
additional coordinate and, in this sense, the system that we are considering
is an extended system. The partition sum for the extended system is 
\begin{equation}
Q(N+1,P,T)=\int_{0}^{1}d\sigma ^{\prime }\ Q(N+1,P,T,\sigma ^{\prime })
\end{equation}
where $Q(N+1,P,T,\sigma ^{\prime })$ is the partition function for the
isothermal-isobaric system with one interstitial particle with radius $%
\sigma ^{\prime }$. The probability of finding the interstitial particle
with a specific radius $\sigma ^{\prime }=\sigma _{I}$ is 
\begin{equation}
P(\sigma _{I}|N+1,P,T)=\frac{\int_{0}^{1}d\sigma ^{\prime }\
Q(N+1,P,T,\sigma ^{\prime })\delta (\sigma _{I}-\sigma ^{\prime })}{Q(N,P,T)}%
=\frac{\ Q(N+1,P,T,\sigma _{I})}{Q(N+1,P,T)}
\end{equation}
and the Gibbs free energy $G(N+1,P,T,\sigma _{I})$ of a system with an
interstitial with diameter $\sigma _{I}$ is equal to 
\begin{equation}
G(N+1,P,T,\sigma _{I})=-k_{B}T\ln Q(N+1,P,T,\sigma _{I})
\end{equation}
Thus, the free energy difference between a system with a pointlike
interstitial ($\sigma _{I}=0$) and a full-grown interstitial ($\sigma _{I}=1$%
) is 
\begin{eqnarray}
g^{grow} & = &G(N+1,P,T,1)-G(N+1,P,T,0)  \nonumber \\
& = &-kT\ln \frac{Q(N+1,P,T,1)}{Q(N+1,P,T,0)}  \nonumber \\
& = &kT\ln \frac{P(0|N+1,P,T)}{P(1|N+1,P,T)}
\end{eqnarray}

It is obvious that $P(\sigma |N+1,P,T)$ is can be very small for large
values of $\sigma _{I}$. In order to get an accurate histogram for $P(\sigma
_{I}|N+1,P,T)$, we have to use a biased sampling scheme. We employ the
method of umbrella-sampling/multicanonical sampling \cite
{Valleau,berg91,smith96}, where we associate a weight $\xi (\sigma )$ with $%
\sigma $, which we use while sampling over $\sigma $: 
\[
P(\sigma |N+1,P,T,\{\xi \})\propto P(\sigma |N+1,P,T)e^{\xi (\sigma )} 
\]
If we sample over this distribution, we get a histogram $P(\{\sigma
\}|N+1,P,T,\{\xi \})$, for which we can get the desired histogram $%
P(\{\sigma \}|N+1,P,T)$ by refolding the bias: 
\[
P(\sigma |N+1,P,T)\propto P(\sigma |N+1,P,T,\{\xi \})e^{-\xi (\sigma )} 
\]

The weights $\xi (\sigma )$ are obtained by iteratively running the system
and calculating (for the run $i+1$, from the results of run $i$) 
\[
\xi _{i+1}(\sigma )=\xi _{i}(\sigma )-\ln P(\sigma |N,P,T,\{\xi _{i}(\sigma
)\})+C 
\]
where $C$ is an arbitrary constant \cite{smith96}. This will make the
histogram $P(\{\sigma _{0}\}|N+1,P,T,\{\xi \})$ converge to a flat
distribution over the accessible range.

\subsection{Interstitial type discrimination}

The fcc crystal has two types of possible places, or `holes' in which
interstitials can reside: one of octahedral shape and one of tetrahedral
shape. There are four octahedral holes and eight tetrahedral holes in one
fcc unit cell. To measure the relative concentrations of interstitials in
these two types of holes, it turns out it is not possible to try to prepare
the system in one hole and calculate $g^{grow}$, because during the course
of a simulation the interstitial makes many hops. These hops are caused by
the interstitial taking the place of a lattice particle which then becomes
the interstitial.

In order to measure the relative occupation probability of the different
holes, the interstitial has to be traced. This is done by the following
scheme: at the start of the simulation, every particle $i$, except the
original interstitial, is assigned to a lattice position ${\bf R}_{i}$. At
fixed sampling intervals, the squared distance between the original
interstitial and the nearest lattice sites ($\delta _{int,i}^{2}=({\bf r}%
_{int}-{\bf R}_{i})^{2}$) is compared with $\delta _{i}^{2}=({\bf r}_{i}-%
{\bf R}_{i})^{2}$. If $\delta _{int,i}^{2}<\delta _{i}^{2}$, the
interstitial and particle $i$ exchange identity (i.e. the interstitial
acquires a lattice position ${\bf R}_{i}$ and particle $i$ becomes the
interstitial). Once we have identified the interstitial, it is
straightforward to assign it to a tetrahedral or octahedral hole.

\section{Simulation results}

\begin{figure}[p]
\center
\includegraphics[angle=0,width=10cm]{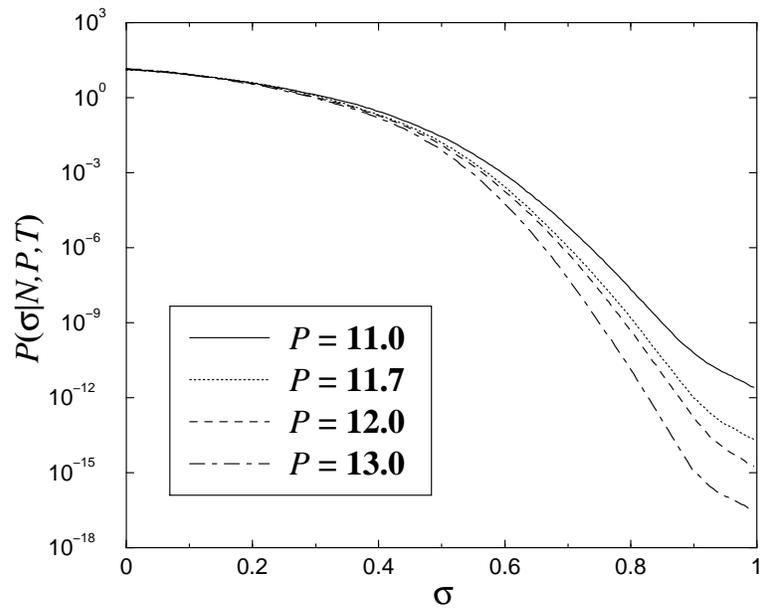}
\caption{The normalized probability $P(\protect\sigma |N,P,T)$ of finding an
interstitial with radius $\protect\sigma /2$ for hard-sphere crystals at
(reduced) pressures 11, 11.7, 12 and 13.  }
\label{pw}
\end{figure}

The free energy calculations were performed with $256+1$ particle systems ($%
4\times 4\times 4$ cubic fcc unit cells) at four different pressures.
Different parts of the  histogram $P(\{\sigma \}|N,P,T,\{\eta \})$ were
calculated in parallel, and subsequently combined.  The calculation of the
weights took about $20$ iterations of $2\cdot 10^{4}$ MC sweeps per CPU on 5
CPU's. Once the weights were known for one pressure, they could be used as
starting points for the other pressures, accelerating the weight calculation
considerably. Final calculations were done with approximately $20$
iterations of $4\cdot 10^{5}$ sweeps each (again on 5 CPU's). The final $%
P(\sigma |N,P,T)$ histograms for all four pressures are plotted in Fig. \ref
{pw}.

\begin{figure}[p]
\center
\includegraphics[angle=270,width=10cm]{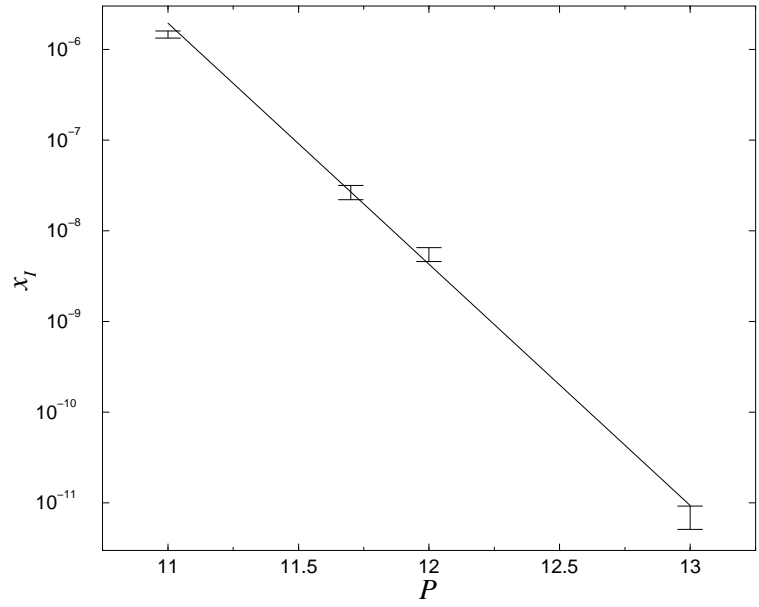}
\caption{$x_{I}$ as a function of  the reduced pressures $P$. The drawn line
corresponds to a fit of the form  $x_{I}=\exp (-6.1P+54)$}
\label{x_fig}
\end{figure}

For the calculation of $\mu $ at the different pressures, the results for
the free energy of the perfect crystal \cite{Polson} were used together with
Hall's\cite{hall70} equation of state. The results are summarized in Fig.~%
\ref{x_fig} and table \ref{restable}. For one pressure ($P=11.7$, the
coexistence pressure), we calculated $g_{I}$ for a larger system ($N=8\times
8\times 8+1=2048+1$) to check for finite-size effects; as can be seen from
the results, these are negligible. Using the interstitial type
discrimination algorithm described above, it was found that the (bigger)
octahedral holes are{\bf \ }far more likely to contain the interstitial than
the tetrahedral holes (see table .\ref{restable}).

\section{Analytical estimate of the free energy of Interstitials}

\label{aest}

As octahedral holes are the largest cavities in a fcc crystal, we limit our
analysis to these. The number of octahedral holes in a fcc crystal is equal
to the number of lattice sites, the derivation of the expression for the
concentration of interstitials is almost identical to the one for the
vacancy concentration. Let us denote the change in free energy associated
with the introduction of an interstitial at a specific octahedral site by $%
f_{I}$. The concentration of interstitials is then 
\begin{equation}
x_{I}=\exp \Big[-\beta (f_{I}-\mu )\Big]
\end{equation}
In a static lattice, $r_{0}$, the radius of such octahedral holes equals $(%
\sqrt{2}/2-0.5)a$, where $a$ is the nearest-neighbor distance. For a
hard-sphere crystal at melting, this radius equals $r_{0}=$ 0.229 $\sigma $.
Clearly, in order to fit in an interstitial, the cavity has to be expanded.
If we assume that the crystal is elastically isotropic (a fair approximation
for a cubic crystal) then the work need to create a cavity of radius $r$
equals\cite{FrenkelKinTheoLiq} 
\begin{equation}
W=8\pi \mu r_{0}(r-r_{0})^{2}  \label{elastic1}
\end{equation}
where $\mu $ is the shear Lam\'{e} coefficient. How large should $r$ be?
Clearly, it should be at least 0.5$\sigma $, otherwise the interstitial
would not fit into the lattice. But, in fact, it should be larger, because
the interstitial particle itself requires some free volume $v_{F}$. We
should therefore minimize the sum of the free energy of a particle in a
cavity of radius $r$ and the elastic energy required to create such a
cavity. Using $v_{F}=(4\pi /3)(r-\sigma /2)^{3}$, the expression for this
free energy is\footnote{%
In what follows, we leave out the factor involving the de Broglie thermal
wavelength, as it cancels anyway in the final result.} 
\begin{equation}
F(r)=-kT\ln \Big[(4\pi /3)(r-\sigma /2)^{3}\Big]+8\pi \mu r_{0}(r-r_{0})^{2}
\label{elastic2}
\end{equation}
Differentiating Eqn. (\ref{elastic2}) with respect to $r$ yields the
following equation for the equilibrium radius of the cavity: 
\begin{equation}
-\frac{3kT(r-\sigma /2)^{2}}{(r-\sigma /2)^{3}}+16\pi \mu r_{0}(r-r_{0})=0
\label{rmin}
\end{equation}
This yields the following equation for $r$%
\[
r^{2}-(\sigma /2+r_{0})r+\frac{\sigma r_{0}}{2}-\frac{3kT}{16\pi \mu r_{0}}=0
\]
and hence 
\begin{equation}
r_{I}=\frac{\sigma /2+r_{0}}{2}+\sqrt{\frac{(\sigma /2+r_{0})^{2}}{4}-\frac{%
\sigma r_{0}}{2}+\frac{3kT}{16\pi \mu r_{0}}}  \label{rcavity}
\end{equation}
Inserting Eqn. (\ref{rcavity}) in Eqn. (\ref{elastic2}) we obtain the
expression for the total free energy of an interstitial at a specific
lattice site 
\begin{equation}
f_{I}=-kT\ln \Big[(4\pi /3)(r_{I}-\sigma /2)^{3}\Big]+8\pi \mu
r_{0}(r_{I}-r_{0})^{2}
\end{equation}
If we use the parameters for a hard-sphere solid at melting ($\mu \approx
C_{44}=46$\cite{FrenkelLaddElastic}), we find that the predicted
concentration of interstitials is approximately $1\cdot 10^{-7}$.
Considering the crudeness of the approximations involved in deriving this
result, the agreement with the corresponding numerical estimate ( $%
x_{I}\approx 3\cdot 10^{-8}$) is gratifying. However, at higher densities,
the agreement becomes worse, possibly because it is no longer justified to
assume isotropic, linear, elastic behavior around an interstitial (see table 
\ref{restable}).

\bigskip 

In summary, we have shown that the equilibrium concentration of
interstitials in hard-sphere crystals is so low that most experiments will
not be able to detect them. We find that interstitials are quite mobile.
This implies that interstitials that are trapped during crystal growth
should be able to diffuse rapidly to the crystal surface. This information
is good news for experimentalists trying to grow photonic bandgap materials.
%initially trapped interstitials are likely to diffuse quickly to the surface 
%where they disappear. 
On the other hand, colloidal hard sphere crystals
will have a high equilibrium concentration of vacancies. 
With the present
accuracy of free-energy calculations, vacancies yield a detectable change in
the free energies, but interstitials do not.

\section{\protect\bigskip Acknowledgment}

The work of the FOM Institute is part of the research program of FOM and was
made possible through financial support by the Dutch Foundation for
Scientific Research (NWO). DF gratefully acknowledges the fact that he first
learned about computer simulation from the famous review by B. J. Berne and
G. D. Harp,''{\em \ ``On the Calculation of Time Correlation Functions''},
Adv. in Chem. Phys. XVII 63, (1970).  

%\bibliography{interstitial}

\newpage 
\begin{table}[p]
%\center

\begin{tabular}{|r|c|c|c|c|c|c|c|c|}
\hline
$P$ & $N$ & $\eta$ & $\mu$ & $g_I/kT$ & $x_I$ & $h_{et}$ & $C_{44}$ & $x_{I,{%
nalytical}}$ \\ \hline
$11.0$ & $256 + 1$ & $0.536$ & $16.5$ & $29.9(1)$ & $1.5 \cdot 10^{-6}$ & $%
0.087(6)$ & $41$ & $1.1\cdot10^{-6}$ \\ 
$11.7$ & $256 + 1$ & $0.545$ & $17.1$ & $34.5(2)$ & $2.7 \cdot 10^{-8} $ & $%
0.032(2)$ & $46$ & $1.3\cdot10^{-7}$ \\ 
$11.7$ & $2048 + 1$ & $0.545$ & $17.1$ & $34.7(2) $ & $2.4 \cdot 10^{-8} $ & 
& $46$ & $1.3\cdot10^{-7}$ \\ 
$12.0$ & $256 + 1$ & $0.548$ & $17.4$ & $36.5(2)$ & $5.6 \cdot 10^{-9} $ & $%
0.079(9)$ & $48$ & $6.8\cdot10^{-8}$ \\ 
$13.0$ & $256 + 1$ & $0.559$ & $18.4$ & $44.1(3)$ & $7.2 \cdot 10^{-12} $ & $%
0.118(8)$ & $57$ & $3.2\cdot10^{-9}$%
\end{tabular}
\caption{Simulation results for the properties of interstitials in hard
sphere crystals.  The values for the packing fraction $\protect\eta $ and
the chemical potential $\protect\mu $ were taken from refs.{\protect\cite
{Polson}} and {\protect\cite{hall70}}. $h_{et}$ is the fraction of
interstitials found in tetrahedral holes. The values from the analytical
estimate of section \ref{aest} are given as $x_{I,{nalytical}}$, using $%
C_{44}$ values interpolated from {\protect \cite{FrenkelLaddElastic}}}
\label{restable}
\end{table}

\newpage 

\newpage 

\end{document}